\documentstyle[sprocl]{article}

\def\ra{\rightarrow}

\def\be{\begin{equation}}
\def\ee{\end{equation}}
\def\bea{\begin{eqnarray}}
\def\eea{\end{eqnarray}}

\begin{document}

\title{NON-EQUILIBRIUM QCD OF HIGH-ENERGY MULTI-GLUON DYNAMICS
\footnote{Based on invited talk given at the 20th Johns Hopkins Workshop on
{\it Non-Perturbative Particle Theory \& Experimental tests},
Heidelberg, Germany, June 27-29, 1996.}
}

\author{Klaus Geiger}
\address{Brookhaven National Laboratory, Upton, N.Y. 11973, U.S.A.}


\maketitle
\abstracts{
I discuss an approach to derive from first principles, a
real-time formalism to study the dynamical 
interplay of quantum and statistical-kinetic properties of 
non-equilibrium multi-parton systems produced in high-energy QCD 
processes. 
The ultimate goal (from which one is still far away) is to have
a practically applicable description of the space-time evolution
of a general  initial system of gluons and quarks, characterized by
some large energy or momentum scale, that expands, diffuses and dissipates
according to the self- and mutual-interactions, and eventually
converts dynamically into final state hadrons.
For example,
the evolution of parton showers in the mechanism
of parton-hadron conversion in high-energy hadronic collisions,
or, the description of
formation, evolution and freezeout  of a quark-gluon plasma,
in ultra-relativistic heavy-ion collisions.}

\section{Introduction}

In general, the study of a high-energy multi-particle system and its
quantum dynamics involves three essential aspects:
first, the aspect of space-time, geometry and
the structure of the vacuum; second, the quantum
field aspect of the particle excitations; and third, the statistical
aspect of their interactions. These three elements are generally
interconnected in a non-trivial way by their overall dynamical
dependence.
Therefore, in order to formulate a  quantum description
of the complex non-equilibrium dynamics, one needs to find a
quantum-statistical and kinetic formulation of QCD that unifies
the three aspects self-consistently.
The main tools to achieve this are:
the {\it closed-time-path} (CTP) formalism 
\cite{schwinger,chou} (for treating initial
value problems of irreversible systems), and (ii)
{\it transport theory} based on Wigner function techniques 
\cite{degroot} 
(for a kinetic description of inhomogenous non-equilibrium systems).

The common feature of high-energy particle collisions is that they allow 
a distinction between
a short-distance quantum field theoretical scale and a 
larger distance statistical-kinetic scale, which is essentially
an effect of ultra-relativistic kinematics.
This advantagous property  facilitates
the passage from {\it exact QCD field theory}
of coherent non-abelian gauge fields to
an {\it approximate quantum kinetic theory} of an ensemble of incoherent
gluons.
When described in a reference frame, in which the particles move close
to the speed of light,
the effects of time dilation and Lorentz contraction
separate the intrinsic quantum motion of the individual
particles from the  statistical correlations among them.
On the one hand,
the quantum dynamics is
determined by the self-interactions of the bare quanta, 
which dresses them  up to quasi-particles with a substructure of quantum fluctuations.
This requires a fully quantum theoretical analysis including renormalization. 
On the other hand, the kinetic dynamics
can well be described statistical-mechanically 
by the motion of the quasi-particles 
that is, 
by binary interactions between these quasi-particles, and
by the possible
presence of a coherent mean color field that may be induced by the
collective motion of the partons.
Such a distinct description of quantum and kinetic dynamics
is possible, because the quantum
fluctuations are highly concentrated around the light cone,
occurring at very short distances, and decouple to very
good approximation from the kinetic evolution
which is dictated by comparably large space-time scales.
As mentioned, 
the natural two-scale separation is just the consequence of time dilation
and Lorentz contraction, and is true for any lightcone
dominated process. In fact, at asymptotic energies the
quantum fluctuations are exactly localized on the lightcone, and
so the decoupling becomes perfect.
This observation is the key to formulate a quantum kinetic description
in terms of particle phase-space densities, involving
a simultanous specification of momentum space and space-time, 
because at sufficiently high energy,
the momentum scale $\Delta p$ of the individual particles' quantum fluctutions
and the scale $\Delta r$ of space-time variations of the system of particles
satisfiy $\Delta p \Delta r \gg 1$, consistent with the uncertainty principle.

In what follows, I am guided by 
the recent paper \cite{ms39} 
and the related literature discussed therein, plus on preliminary results
of work in progress \cite{ms42}.
For purpose of lucidity, I will henceforth confine myself to 
pure Yang-Mills theory, i.e.
consider gluons only and ignore the quark degrees of freedom. The latter are
straightforward to include.

\section{Non-equilibrium techniques for QCD}

\subsection{Basics of the closed-time-path formalism}

As proclaimed, the goal is  to describe the time evolution of
a  non-equilibrium quantum system consisting of an initial ensemble
of  high-energy gluons at starting time $t_0$.
In this context, the starting point of non-equilibrium
field theory  is to write down the CTP {\it in-in amplitude} $Z_P$ 
for the evolution of
the initial quantum state $\vert in\rangle$ forward in time
into the remote future, in the presence of 
a medium which described by the density matrix. The amplitude
$Z_P$ is formally given by \cite{chou}:
\begin{equation}
Z_P[{\cal J},\hat{\rho}] 
\;=\;
\langle \; in \;\vert \;in\;\rangle _{{\cal J}}
\,\;=\,\;
\mbox{Tr}\left\{
U^\dagger(t_0 ,t)\, U(t,t_0)\, \hat{\rho}(t_0)\right\}
_{{\cal J}}
\;,
\label{Z1}
\end{equation}
where 
${\cal J} = ({\cal J}^+,{\cal J}^-)$ is an external source with components on the $+$ and
$-$ time branch.
$\hat{\rho}(t_0)$ denotes the 
The initial state  density matrix is denoted $\hat{\rho}(t_0)$,
$U$ is the time evolution operator, and $T$ ($T^\dagger$) 
denotes the time (anti-time) ordering.
Within the CTP formalism the amplitude $Z_P$ can be evaluated by time integration
over the {\it closed-time-path} $P$ in the complex $t$-plane.
This closed-time path 
extends from $t=t_0$ to $t=t_\infty$ in the remote future
along the positive ($+$) branch and back to $t=t_0$ along the negative ($-$) branch.
where any point on the $+$ branch is understood at an earlier instant
than any point on the $-$ branch.
With $Z_P$ defined on this closed-time-path, 
one may then, as in standard field theory, derive from it 
the Green functions and their equations of motion. The differences
between the CTP and the standard field theory, which are briefly
summarized below,  arise then solely from
the different time contour.
\smallskip

The interpretation of this formal apparatus for the
evolution along the closed-time path $P$ is rather simple:
If the initial state is the vacuum itself, that is, 
the absence of a medium generated
by other particles, then the density matrix $\hat{\rho}$ is diagonal and
in (\ref{Z1}), one has  $\vert in\rangle \rightarrow \vert 0 \rangle$. 
In this case the evolution along the $+$ branch is identical to the anti-time ordered
evolution along the $-$ branch (modulo an irrelevant phase),
and space-time points on different branches cannot cross-talk.
In the presence of a medium however, the density matrix contains off-diagonal elements,
and there are statistical correlations between the quantum system and the
medium particles (e.g. scatterings) that lead to correlations between space-time 
points on the $+$ branch with space-time points on the $-$ branch.
Hence, when addressing the evolution of a multi-particle system, both the
deterministic self-interaction of the quanta, i.e. 
the time (anti-time) ordered evolution along the $+$ ($-$) branch,
{\it and} the statistical mutual interaction with each other,
i.e. the cross-talk between $+$ and $-$ branch,  
must be included in a self-constistent manner. 
The CTP method achieves this through the time integtation along the contour $P$.
Although for physical observables the time values are on the $+$ branch, 
both $+$ and $-$ branches will come into play at intermediate steps in 
a self-consistent calculation.
\smallskip

The convenient feature of this Green function formalism on the closed-time path 
is that it is formally completely analogous to standard quantum field theory,
except for the fact that the fields have contributions from both
time branches, and the path-integral representation of the {\it in-in} amplitude
(\ref{Z1}), contains as usual the classical action $I[{\cal A}$ and
source terms ${\cal J}\circ{\cal A}$, but now for both time branches,
\newpage
\begin{eqnarray}
Z_P[{\cal J}^+,{\cal J}^-,\hat{\rho}]
&=&
\int \,{\cal D} {\cal A}^+{\cal D} {\cal A}^-
\;\, \exp \left[i \left( \frac{}{} I[{\cal A}^+] \,+\, {\cal J}^+\cdot{\cal A}^+
\right)
\right.
\label{Z14a}
 \\
& &
\;\;\;\;\;\;\;\;\;\;\;\;\;\;\;\;\;\;\;\;\;\;\;\;\;\;\;\;\;\;\;\;
\left.
\;-\;  
i \left( \frac{}{} I^\ast[{\cal A}^-] \,+\, {\cal J}^-\cdot{\cal A}^-
\right)
\right]
\;{\cal M}[\hat{\rho}]
\nonumber
\;.
\end{eqnarray}
From this path-integral representation one obtains 
the  $n$-point Green functions $G^{(n)}(x_1,\ldots, x_n)$, which
are now
$$
G^{\alpha_1\alpha_2 \ldots\,\alpha_n}(x_1,\ldots, x_n)
\;\;, \;\;\;\;\;\;\;\;\;\;\;\;
\alpha_i \;= \pm
\;,
$$
depending on whether the space-time points
$x_i$ lie on the $+$ or $-$ time branch, and
it is possible to construct
a perturbative expansion of the non-equilibrium Green functions in terms
of modified Feynman rules (as compared to standard field theory),
\begin{description}
\item[(i)]
The number of elementary vertices is doubled,
because each propagator line of a Feynman diagram can be
either of the four components of the Green functions.
The  interaction  vertices in which 
all the fields are on the $+$ branch are the usual ones,
while the vertices in which the fields are on 
the $-$ branch have the opposite sign.
On the other hand, combinatoric factors, rules for loop integrals, etc.,
remain exactly the same as in usual field theory.
\smallskip

\item[(ii)]
All local 1-point functions, such as the 
gauge-field or the color current, are `vectors' with 2 components,
\begin{equation}
{\cal A}(x) \;\equiv\;
\left( \begin{array}{c}
{\cal A}^{+} \\ {\cal A}^{-}
\end{array}\right) 
\;\;\;\;\;\;\;\;\;\;\;\;\;\;\;\;
{\cal J}(x) \;\equiv\;
\left( \begin{array}{c}
{\cal J}^{+} \\ {\cal J}^{-}
\end{array}\right) 
\end{equation}
Similarly, all 2-point functions, as
the  gluon propagator $\Delta_{\mu\nu}$ and the polarization tensor 
$\Pi_{\mu\nu}$, are  2$\times$2 matrices with components
\begin{equation}
\Delta(x_1,x_2) \;\equiv\;
\left(\begin{array}{cc}
\Delta^{++} & \;\Delta^{+-} \\
\Delta^{-+} & \;\Delta^{--}
\end{array} \right) 
\;\;\;\;\;\;\;\;
\Pi(x_1,x_2) \;\equiv\;
\left(\begin{array}{cc}
\Pi^{++} &  \;\Pi^{+-} \\
\Pi^{-+} &  \;\Pi^{--}
\end{array} \right) 
\;.
\nonumber
\end{equation}
Explicitely, the components of the propagator are 
\begin{eqnarray}
\Delta_{\mu\nu}^{F}(x,y)\;&\equiv&
\Delta_{\mu\nu}^{++}(x,y)\;=\;
-i\,
\langle \;T\, {\cal A}_\mu^+(x)\, {\cal A}_\nu^+(y) \;\rangle
\nonumber \\
\Delta_{\mu\nu}^{<}(x,y) \;&\equiv&
\Delta_{\mu\nu}^{+-}(x,y)\;=\;
-i\,
\langle \; {\cal A}_\nu^+(y)\, {\cal A}_\mu^-(x) \;\rangle
\nonumber \\
\Delta_{\mu\nu}^{>}(x,y)\;&\equiv&
\Delta_{\mu\nu}^{-+}(x,y)
\;=\;
-i\,
\langle \; {\cal A}_\mu^-(x)\, {\cal A}_\nu^+(y) \;\rangle
\nonumber\\ 
\Delta_{\mu\nu}^{\overline{F}}(x,y) \;& \equiv&
\Delta_{\mu\nu}^{--}(x,y)\;=\;
-i\,
\langle \;\overline{T}\, {\cal A}_\mu^-(x)\, {\cal A}_\nu^-(y) \;\rangle
\label{D22}
\;,
\end{eqnarray}
where
$\Delta^F$ is the usual time-ordered Feynman propagator, $\Delta^{\overline{F}}$
is the corresponding anti-time-ordered propagator, and $\Delta^>$ ($\Delta^<$) is
the unordered correlation function for $x_0 > y_0$ ($x_0 < y_0$).
In compact  notation,
\begin{equation}
\Delta_{\mu\nu}(x,y)\;=\;
-i\,\langle \,T_P {\cal A}(x) {\cal A}(y) (y) \,\rangle
\; ,
\end{equation}
where
the generalized time-ordering operator $T_P$ is defined as
\begin{equation}
 T_P\,A(x)B(y)
\;:=\;
\theta_P (x_0,y_0)\, A(x) B(y) \;+ \;\theta_P(y_0,x_0) \,B(y) A(x)
\label{gto1}
\;, 
\end{equation}
with the $\theta_P$-function  defined as 
\begin{equation}
\theta_P(x_0,y_0)\;=\;
\left\{
\begin{array}{ll}
1 & \;\; \mbox {if $x_0$ {\it succeeds} $y_0$ on the contour $P$}
\\
0 & \;\;  \mbox{if $x_0$ {\it precedes} $y_0$ on  the contour $P$}
\end{array}
\right. 
\;.
\label{gto2}
\end{equation}
Higher order products $A(x)B(y)C(z)\ldots$ are ordered analogously.
Finally, for later use, let me also introduce
the generalized $\delta_P$-function defined on the closed-time path $P$:
\begin{equation}
\delta^4_P(x,y) \;\,:=\;\,
\left\{
\begin{array}{ll}
+\delta^4(x-y) & \;\; \mbox {if} \; x_0 \;\mbox{and} \;y_0 \; \mbox{from positive
branch} \\
-\delta^4(x-y) & \;\; \mbox {if} \; x_0 \;\mbox{and} \;y_0 \; \mbox{from negative
branch} \\
0 & \;\; \mbox {otherwise}
\end{array}
\right. 
\;.
\label{gto3}
\end{equation}
\end{description}

Henceforth
I will not explicitly label the $+$, $-$ components,
unless it is necessary. Instead a compressed notation 
is used, in which it is understood that, e.g.,  1-point functions
such as ${\cal A}(x)$ or ${\cal J}(x)$, 2-point functions 
such as $\Delta_{\mu}(x,y)$
or $\Pi_{\mu\nu}(x,y)$, receive contributions from both $+$ and $-$ time 
branches.

\subsection{The generating functional for the non-equilibrium Green functions}

The amplitude $Z_P$ introduced in (\ref{Z1}) admits a path-integral representation which
gives the  {\it generating functional for the CTP Green functions}
defined on closed-time-path $P$ \cite{ms42}:
\begin{equation}
Z_P[{\cal J},\hat{\rho}] \;=\;
{\cal N}\;\int \,{\cal D}{\cal A} \; \mbox{det}{\cal F}\;\delta \left( f[{\cal A}]\right)
\;\,
\exp \left\{i \left( \frac{}{} I \left[ {\cal A}, {\cal J}\right]
\right)\right\}
\;\; {\cal M}(\hat{\rho})
\label{Z2}
\;,
\end{equation}
where ${\cal A} = ({\cal A}^+,{\cal A}^-)$ and  
${\cal J} = ({\cal J}^+,{\cal J}^-)$ 
have two components, living on the $+$ and $-$ time branches.
\medskip

\noindent
The structure of the functional $Z_P$ in (\ref{Z2}) 
is the following:

\begin{description}
\item[(i)]
The functional integral (with normalization ${\cal N}$) 
is over all gauge field configurations
with measure ${\cal D}{\cal A} \equiv \prod_{\mu,a} {\cal D}{\cal A}_\mu^a$,
subject to the condition of gauge fixing, here  for the 
{\it class of non-covariant gauges} defined by
\begin{equation}
f^a[{\cal A}] \;:=\; \hat{n}\cdot {\cal A}^a(x) \;-\; B^a(x)
\;\;\;\;\;\;\;\;\;\;\;
\Longrightarrow
\;\;\;\;\;\;\;\;\;\;
\langle \;\hat{n}^\mu\,{\cal A}_\mu^a(x)\;\rangle \;=\;0
\label{gauge1}
\;,
\end{equation}
where
$\hat{n}^\mu\,\equiv \,\frac{n^\mu}{\sqrt{\vert n^2\vert }}$ and
$n^\mu$ is a constant 4-vector, being either space-like ($n^2 < 0$), 
time-like ($n^2 > 0$), or light-like ($n^2=0$).
With this choice of gauge class the {\it local gauge constraint}
on the fields ${\cal A}^a_\mu(x)$ in
the path-integral (\ref{Z2}) becomes,
\begin{eqnarray}
\mbox{det}{\cal F}\;\delta \left( \hat{n}\cdot {\cal A}^a-B^a\right)
&=&
\mbox{const}\,\times\,
\exp\left\{- \frac{i}{2\alpha}
\,\int_P d^4x \,\left[\hat{n}\cdot {\cal A}^a(x)\right]^2
\right\}
\nonumber \\
&\equiv&
I_{GF}\left[\hat{n}\cdot{\cal A}\right]
\;,
\label{gauge2}
\end{eqnarray}
where $\mbox{det}{\cal F}$ is the Fadeev-Popov determinant (which in the case
of the non-covariant gauges turns out to be a constant factor),
and where $\delta (\hat{n}\cdot{\cal A}) \equiv \prod_{a}\delta (\hat{n}\cdot{\cal A}^a)$.
The right side translates this constraint into a
the {\it gauge fixing} functional $I_{GF}$.
The particular choice of 
the vector $\hat{n}^\mu$ and of 
the real-valued parameter $\alpha$ is dictated by
the physics or computational convenience, and distinguishes 
further within the class
of non-covariant gauges \cite{gaugereview,gaugebook}:
\begin{eqnarray}
\mbox{homogenous axial gauge}: \;& & \;\;\;\;\; n^2 \;< \; 0 \;\;\;\;\;\;\alpha \;=\;0 
\nonumber \\
\mbox{inhomogenous axial gauge}: \;& & \;\;\;\;\; n^2 \;< \; 0 \;\;\;\;\;\;\alpha \;=\;1 
\nonumber \\
\mbox{temporal axial gauge}: \;& & \;\;\;\;\; n^2 \;> \; 0 \;\;\;\;\;\;\alpha \;=\;0 
\nonumber \\
\mbox{lightcone gauge}: \;& & \;\;\;\;\; n^2 \;= \; 0 \;\;\;\;\;\;\alpha \;=\;0 
\label{gauge1a}
\;.
\end{eqnarray}

\item[(ii)]
The exponential $I$ is the {\it effective classical action} 
with respect to both the $+$ and the $-$ time contour, 
$
I\left[ {\cal A}, {\cal J}\right]
\equiv I\left[ {\cal A}^+, {\cal J}^+\right]
\;-\;
I^\ast\left[ {\cal A}^-, {\cal J}^-\right]
$,
including
the usual Yang-Mills action $I_{YM}=\int d^4x {\cal L}_{YM}$, plus the source
${\cal J}$ coupled to the gauge field ${\cal A}$:
\begin{eqnarray}
I\left[ {\cal A}, {\cal J}\right]
& =&
-\frac{1}{4} \int_P d^4x \,{\cal F}_{\mu\nu}^a(x) {\cal F}^{\mu\nu, \,a}(x) 
\;+\;
\int_P d^4x \,{\cal J}_{\mu}^a(x) {\cal A}^{\mu, \,a}(x) 
\nonumber \\
& &\nonumber \\
&\equiv&
\;\;\;\;\;\;\;\;
I_{YM}\left[{\cal A}\right] \;\;\;+\;\;\; {\cal J}\circ {\cal A}
\label{I1}
\;,
\end{eqnarray}
where
$
{\cal F}_{\mu\nu}^a=
\partial^x_\mu {\cal A}_{\nu}^a - \partial^x_\nu {\cal A}_{\mu}^a 
+ g\, f^{abc} \,{\cal A}_\mu^b {\cal A}_\nu^c
$.

\item[(iii)]
The form of the initial state at $t=t_0$ as
described by the density matrix $\hat{\rho}$ is
embodied in the function      
${\cal M}(\hat{\rho})$ which is the density-matrix element of the gauge fields
at initial time $t_0$, 
\begin{equation}
{\cal M}(\hat{\rho}) 
\;=\;
\langle \,{\cal A}^+ (t_0) \vert \,{\hat \rho}
\,\vert\,{\cal A}^-(t_0)\,\rangle
\;\equiv\;\,
\exp\left( i\; {\cal K}[{\cal A}] \right)
\label{K}
\;,
\end{equation}
where ${\cal A}^\pm$ refers to the $+$ and $-$ time branch {\it at} $t_0$,
respectively.
The functional ${\cal K}$ may be expanded in a series of
non-local kernels corresponding to multi-point correlations concentrated
at $t=t_0$,
\begin{eqnarray} 
{\cal K}[{\cal A}]
&=&
{\cal K}^{(0)} \;+\;
\int_P d^4x \;{\cal K}^{(1)\;a}_{\;\;\;\,\mu}(x) \;{\cal A}^{\mu, \,a}(x) 
\nonumber \\
& &
\;\;\;\;\;\;\;
\;+\;
\frac{1}{2}
\int_P d^4xd^4y \;{\cal K}^{(2) \;ab}_{\;\;\;\,\mu\nu}(x,y)\;{\cal A}^{\mu, \,a} (x)
\,{\cal A}^{\nu, \,b}(y) 
\,\ldots
\label{Kexpansion}
\;.
\end{eqnarray}
Clearly, the sequence of kernels ${\cal K}^{(n)}$ contains as much information
as the original density matrix.
In the special case that $\hat{\rho}$ is diagonal,
the kernels  ${\cal K}^{(n)}=0$ for all $n$, and 
the  usual `vacuum field theory' is recovered.
\end{description}
\medskip

The path-integral representation (\ref{Z2}) can be rewritten in a 
form more convenient for the following:
First, the gauge-fixing functional $I_{GF}[\hat{n}\cdot{\cal A}]$ is implemented by
using (\ref{gauge2}).
Second,  the series representation (\ref{Kexpansion}) is inserted into
the initial state functional ${\cal M}(\hat{\rho})$.
Third,  
${\cal K}^{(0)}$ is absorbed in the overall normalization ${\cal N}$ of $Z_P$ 
(henceforth set to unity),
and  the external source ${\cal J}$ in the 1-point kernel ${\cal K}^{(1)}$:
\begin{equation}
{\cal K}^{(0)}\;:=\; i\, \ln {\cal N}
\;,\;\;\;\;\;\;\;\;\;\;\;\;\;\;
{\cal K}^{(1)} \;:=\; {\cal K}^{(1)}\;+\; {\cal J} 
\label{NJ}
\;.
\end{equation}
Then  (\ref{Z2}) becomes,
\begin{equation}
Z_P[{\cal J},\hat{\rho}] 
\;\;\Longrightarrow\;\;
Z_P[{\cal K}] \;=\;
\int \,{\cal D}{\cal A} \; 
\exp \left\{i \left( \frac{}{} I \left[ {\cal A}, {\cal K}\right]
\right)\right\}
\label{Z3}
\;,
\end{equation}
where, instead of (\ref{I1}),
\begin{equation}
I \left[ {\cal A}, {\cal K}\right]
\;\equiv\;
I_{YM}\left[{\cal A}\right] 
\;+\;
I_{GF}\left[\hat{n}\cdot{\cal A}\right]
\;+\; {\cal K}^{(1)}\circ {\cal A} \;+\; 
\frac{1}{2}\,{\cal K}^{(2)}\circ \left({\cal A}\,{\cal A}\right)
\;+\;
\frac{1}{6}\,{\cal K}^{(3)}\circ \left({\cal A}\,{\cal A} \,{\cal A}\right)
\;+\;\ldots
\label{I2}
\;.
\end{equation}
\medskip

\section{Separating soft and hard dynamics and the equations of motion}

The first step in the strategy is a separation of
soft and hard physics in the
path-integral formalism with Green functions of
both the soft and hard quanta in the presence of the soft classical field
is induced by and feeding back to the quantum dynamics.
The basic idea to split up the gauge field
${\cal A}_\mu$ appearing in the classical action 
$I_{YM}\left[{\cal A}\right]$ into a soft (long-range) part $A_\mu$,  
and a hard (short-range) quantum field $a_\mu$:
\begin{eqnarray}
{\cal A}_\mu^a(x) &=&
\int \frac{d^4k}{(2\pi)^4}\,e^{+i k \cdot x} 
{\cal A}_\mu^a(k) \;\, \theta(\mu - k^0) 
\label{Aa}
\\
& &
\;\;\;\;\;\;\;\;\;
\;\,+\;\,
\int \frac{d^4k}{(2\pi)^4}\,e^{+i k \cdot x} 
{\cal A}_\mu^a(k) \,\; \theta(k^0 -\mu)
\;\;\equiv\;\;
A_\mu^a(x) \;+\; a_\mu^a(x)
\;.
\nonumber
\end{eqnarray}
This is the formal definition of the terms `soft' and `hard'.
The soft and hard physics are separated by a
(at this point arbitrary) space-time scale $\lambda \equiv 1/\mu$,
so that one may associate 
the soft field $A_\mu$ being responsible for long range color collective effects, 
and the hard field $a_\mu$ embodying the short-range quantum dynamics.
Consequently, the field strength tensor receives a soft,
a hard part, a mixed contribution,
\begin{equation}
{\cal F}_{\mu\nu}^{a}(x) \;\equiv\;
\left(\frac{}{}
F_{\mu\nu}^{a}[A] \;+\; f_{\mu\nu}^{a}[a] \;+\; \phi_{\mu\nu}^{a}[A,a]
\right) (x)
\label{Ffphi}
\;.
\end{equation}

Now comes physics input. Consider the following
{\it physics scenario}: The initial state is
a (dilute) ensemble of hard gluons of very small spatial extent
$\ll \lambda$, corresponding to transverse momenta $k_\perp^2 \gg \mu^2$.
By definition of $\lambda$, or $\mu$, the short-range character of
these quantum fluctuations implies that the expectation value $\langle a_\mu\rangle$
vanishes at all times. However, the long-range correlations of the eventually
populated soft modes with very small momenta $k_\perp^2\ll\mu^2$ may lead
to a collective mean field with non-vanishing $\langle A\rangle$.
Accordingly,  the following condition
on the expectation values of the fields is imposed:
\begin{equation}
\langle \; A_\mu^a(x)\;\rangle \;
\left\{
\begin{array}{ll}
\;=\;0 \;\;\;\mbox{for} \;t\,\le\,t_0 \\
\;\ge\;0 \;\;\;\mbox{for} \;t\,>\,t_0
\end{array}
\right.
\;\;\;\;\;\;\;\;\;\;\;\;\;\;\;\;\;\;\;\;
\langle \; a_\mu^a(x)\;\rangle \;\stackrel{!}{=}\; 0
\;\;\;\mbox{for all} \;t
\;.
\label{MFconstraint}
\end{equation}
Furthermore, for simplicity the quantum fluctuations of the soft field are ignored, assuming
any multi-point correlations of soft fields to be small,
\begin{equation}
\langle \; A_{\mu_1}^{a_1}(x_1)\; \ldots A_{\mu_n}^{a_n}(x_n) \;\rangle
\ll\;\langle \;  A_{\mu_1}^{a_1}(x_1)\;\rangle
\;\ldots \;\langle \;  A_{\mu_1}^{a_n}(x_n)\;\rangle
\;,
\end{equation}
i.e. take $A_\mu$ as a non-propagating and non-fluctuating, classical field.

When quantizing this decomposed theory by writing down the
appropriate {\it in-in}-amplitude $Z_P$, one must be
consistent with the gauge field decomposition (\ref{Aa}) into soft and 
hard components and with the classical character of the former.
$M^{(1)}_{\mu} = 0$, $M^{(2)}_{\mu\nu}\ge 0$. That is, I restrict
in the following to a class of non-equilibrium initial states of
Gaussian form (i.e. quadratic in the $a_\mu$ fields) and
do not consider possible linear force terms.

Substituting the soft-hard mode decomposition (\ref{Aa}) 
with the condition (\ref{MFconstraint}) into (\ref{Z3}), 
the functional integral of the {\it in-in} amplitude (\ref{Z3}) becomes:
\begin{equation}
Z_P[ {\cal  K}] \;=\;
\int \,{\cal D} A \,{\cal D} a \; 
\exp \left\{i \left( \frac{}{} I \left[ A\right]
\;+\; I \left[ a\right] \;+\; I \left[ A, a\right]
\right)\right\}
\label{Z4}
\;,
\end{equation}
with a soft, hard, and mixed contribution, respectively \cite{ms42}.

Introducing the {\it connected} generating functional for the 
{\it connected} Green functions, denoted by ${\cal G}^{(n)}$,
\begin{equation}
W_P\left[{\cal K}\right] \;\;=\;\;
- i \,\ln\, Z_P\left[ {\cal K}\right]
\label{W0}
\;,
\end{equation}
from which one obtains
the {\it connected} Green functions ${\cal G}^{(n)}$
by functional differentiation,
in terms of mixed products of $a_\mu$ and $A_\mu$ fields
\begin{equation}
(-i)\, {\cal G}_{\;\;\;\;\mu_1\ldots \mu_n}^{(n)\;a_1\ldots a_n}(x_1,\ldots, x_n)
\;\equiv\;
\left.
\frac{\delta}{i \,\delta{\cal K}^{(n)}}
W_P[{\cal K}]\right|_{{\cal K}=0}
\label{Green2}
\;,
\end{equation}
where the superscript $(c)$ indicates the `connected parts'.
Specifically, one finds
\begin{eqnarray}
{\cal G}_{\;\;\;\;\mu}^{(1)\;a}(x)
&=&
\langle \;A_\mu^a(x) \;\ \rangle_P^{(c)}
\;\;\equiv\;\; \overline{A}_\mu^a(x)
\nonumber \\
{\cal G}_{\;\;\;\;\mu\nu}^{(2)\;ab}(x,y)
&=&
\langle \;a_\mu^a(x) a_\nu^b(y)\;\rangle_P^{(c)}
\;\;\equiv\;\; 
i \widehat{\Delta}_{\mu\nu}^{ab}(x,y)
\;.
\label{Green3}
\end{eqnarray}
These relations 
define the soft mean field $\overline{A}$ and the hard propagators 
$\widehat{\Delta}$.
\medskip

The equations of motions for $\overline{A}$ and $\widehat{\Delta}$ follow
now as in usual field theory by functional differentiation of the 
effective action, 
\begin{equation}
\Gamma_P\left[\overline{A}, \widehat{\Delta}\right]
\;=\;
W_P\left[{\cal K}\right] 
\;-\;{\cal K}^{(1)}\circ \overline{A}
\;-\;\frac{1}{2}{\cal K}^{(2)}\circ
\left( i\widehat{\Delta}+ \overline{A}^2\right)
\;.
\end{equation}
Note that the main approximation at this point
is  the truncation of the infinite hierarchy of
equations for the $n$-point Green functions of the excact theory,
to the 1-point function (the soft mean field $\overline{A}(x)$) and the
2-point function (the hard propagator $\widehat{\Delta}(x,y)$),
with all higher-point functions being combinations of these and connected
by the 3-gluon and 4-gluon vertices.

\subsection{Yang-Mills equation for the `soft' mean field}

The equation of motion for the soft field $\overline{A}_\mu^a(x)$,
is given by
$
\delta \Gamma_P
/
\delta \overline{A}
=
-{\cal K}^{(1)}
- {\cal K}^{(2)} \circ\overline{A}
$,
from which one obtains,
upon taking into account the initial condition
${\cal K}^{(1)} = 0$, the {\it Yang-Mills equation for} $\overline{A}$:
\begin{equation}
\left[\frac{}{}
\overline{D}^{\lambda,\;ab}  , \; \overline{F}_{\lambda \mu}^{b}
\right](x)
\;=\;
- \,\widehat{j}_{\mu}^{a}(x) 
\;-\; \int_P d^4y \,{\cal K}_{\;\;\;\mu\lambda}^{(2)\,ab}(x,y)
\,\overline{A}^{\lambda,\,b}(y)
\label{YME2}
\;,
\end{equation}
where $[\overline{D}, \overline{F}] =\overline{D}\, \overline{F} -
\overline{F}\, \overline{D}$
with the covariant derivative defined as
$\overline{D}^\lambda \equiv D^\lambda[\overline{A}] = 
\partial_x^\lambda - ig \overline{A}^\lambda$,
and $\overline{F}_{\lambda \mu}\equiv F_{\lambda \mu}[\overline{A}]
= \left[ \overline{D}_{\lambda}\,,\, \overline{D}_{\mu}\right]/(-ig)$.
The left hand side of (\ref{YME2}) may be written as
\begin{equation}
\left[\frac{}{}
\overline{D}^{\lambda,\;ab}  , \; \overline{F}_{\lambda \mu}^{b}
\right](x)
\;=\;
{\cal D}_{(0)\;\mu\nu}^{-1\;\;ab} \;\overline{A}^{\lambda,\,b}(x)
\;\;+\;\;\overline{\Xi}_{\mu}^{a}(x)
\end{equation}
\begin{equation}
{\cal D}_{(0)\;\mu\nu}^{-1\;\;ab} 
\;\equiv\;\delta^{ab} \,\left( g_{\mu\lambda} \partial_x^{2} - 
\partial^x_\mu\partial^x_\lambda 
- \hat{n}_\mu\hat{n}_\lambda \right)
\;,
\label{DF1}
\end{equation}
where, upon taking into account the gauge constraint (\ref{gauge1}),
the $-\hat{n}_\mu\hat{n}_\lambda \overline{A}^\lambda$ 
does not contribute, because
$0=\langle \hat{n}\cdot A\rangle = \hat{n}^\nu \overline{A}_\nu$,
and where
\begin{equation}
\overline{\Xi}_{\mu}^{a}(x)
\;\;=\;\;
\overline{\Xi}_{(1)\;\mu}^{a}(x)\;\;+\;\; \overline{\Xi}_{(2)\;\mu}^{a}(x)
\label{DF2} 
\end{equation}
\begin{eqnarray}
\overline{\Xi}_{(1)\;\mu}^{a}(x)
&= &
\;-\; \frac{g}{2} \,
\int_P\prod_{i=1}^{2} d^4x_i \;
V_{(0)\;\mu\nu\lambda}^{\;\;\;\;abc}(x,x_1,x_2) 
\;\overline{A}^{\nu,\,b}(x_1) \overline{A}^{\lambda,\,c}(x_2)
\label{DF3} \\
\overline{\Xi}_{(2)\;\mu}^{a}(x)
&= &
\;+\;
\frac{i\,g^2}{6} \,
\int_P\prod_{i=1}^{3} d^4x_i \;
W_{(0)\;\mu\nu\lambda\sigma}^{\;\;\;\;abcd}(x,x_1,x_2,x_3) 
\nonumber \\
& & \;\;\;\;\;\;\;\;\;\;\;\;\;\;
\;\;\;\;\;\;\;\;\;\;\;
\times
\;\overline{A}^{\nu,\,b}(x_1) \overline{A}^{\lambda,\,c}(x_2)\overline{A}^{\sigma,\,d}(x_3)
\label{DF4}
\;.
\end{eqnarray}
On the right hand side of (\ref{YME2}), the current $\widehat{j}$ is the
{\it induced current} due to the `hard' quantum dynamics
in the presence of the `soft' field $\overline{A}$:
\begin{equation}
\widehat{j}_{\mu}^{a}(x) 
\;\;=\;\;
\widehat{j}_{(1)\;\mu}^{a}(x) \;\;+\;\;\widehat{j}_{(2)\;\mu}^{a}(x) 
\;\;+\;\;\widehat{j}_{(3)\;\mu}^{a}(x)
\label{J}
\end{equation}
\begin{eqnarray}
\widehat{j}_{(1)\;\mu}^{a}(x) 
& = &
-\; \frac{i\,g}{2} \,
\int_P\prod_{i=1}^{2} d^4x_i \;
V_{(0)\;\mu\nu\lambda}^{\;\;\;\;abc}(x,x_1,x_2) 
\;\widehat{\Delta}^{\nu\lambda,\,bc}(x_1,x_2)
\label{J1}
 \\
\widehat{j}_{(2)\;\mu}^{a}(x) 
&= &
-\;
\frac{g^2}{2} \,
\; \int_P\prod_{i=1}^{3} d^4x_i
\;\;W_{(0)\;\mu\nu\lambda\sigma}^{\;\;\;\;abcd}(x,x_1,x_2,x_3)\;
\nonumber \\
& & \;\;\;\;\;\;\;\;\;\;\;\;\;\;
\;\;\;\;\;\;\;\;\;\;\;
\times
\;\overline{A}^{\nu,\,b}(x_1)
\;\widehat{\Delta}^{\lambda\sigma,\,cd}(x_2,x_3)
\label{J2}
\\
\widehat{j}_{(3)\;\mu}^{a}(x) 
&= &
-\;
\frac{i g^3}{6} \,
\; \int_P\prod_{i=1}^{3} d^4x_i d^4y_i
\;\;W_{(0)\;\mu\nu\lambda\sigma}^{\;\;\;\;abcd}(x,x_1,x_2,x_3)\;
\widehat{\Delta}^{\nu\nu',\,bb'}(x_1,y_1) \;
\nonumber \\
& &
\times \;
\widehat{\Delta}^{\lambda\lambda',\,cc'}(x_2,y_2) \;
\widehat{\Delta}^{\sigma\sigma',\,dd'}(x_3,y_3) \;
\;V_{(0)\;\mu'\nu'\lambda'\sigma'}^{\;\;\;\;abcd}(y_1,y_2,y_3)
\label{J3}
\;.
\end{eqnarray}
Finally, the second term on the right side of (\ref{YME2}) is the
initial state contribution to the current, which vanishes for 
$t = x^0 > t_0$.
\smallskip

Notice that the function $\overline{\Xi}$ on the left hand side of (\ref{YME2})
contains the non-linear self-coupling of the soft field $\overline{A}$
alone, whereas the induced current $\widehat{j}$ on the right hand side
is  determined by the hard propagator $\widehat{\Delta}$,
thereby generating the soft field.

\subsection{Dyson-Schwinger equation for the `hard' Green function}

The  equation of motionfor the `hard' propagator, 
$\widehat{\Delta}_{\mu\nu}^{ab}(x,y)$,  is 
$
\delta \Gamma_P
/
\delta  \widehat{\Delta} 
=
{\cal K}^{(2)}/(2i)
$,
from which one finds after incorporating  the initial condition ${\cal K}^{(1)}=0$,
the {\it Dyson-Schwinger equation for} $\widehat{\Delta}$:
\begin{equation}
\left[\frac{}{}
\left(\widehat{\Delta}_{\mu\nu}^{ab}\right)^{-1} \;-\; 
\left(\Delta_{(0)\,\mu\nu}^{\;\;ab}\right)^{-1} \;+\; 
\overline{\Pi}_{\mu\nu}^{ab} \;+\;\widehat{\Pi}_{\mu\nu}^{ab} 
\,\; \right](x,y)
\;=\;
{\cal K}^{(2)\;ab}_{\;\;\;\;\mu\nu}(x,y)
\label{DSE2}
\;,
\end{equation}
where
$\widehat{\Delta} \equiv \widehat{\Delta}_{[\overline{A}]}$ 
is the {\it fully dressed propagator} of
the `hard' quantum fluctuations in the presence of the `soft' mean field,
whereas $\Delta_{(0)}$ is the {\it free propagator}.
The polarization tensor $\Pi$ has been decomposed in two parts,
a mean-field part, and a quantum fluctuation part.
The {\it mean-field polarization tensor} $\overline{\Pi}$ incorporates
the {\it local} interaction between the `hard' quanta and the `soft' mean field, 
\begin{equation}
\overline{\Pi}_{\mu\nu}^{ab}(x,y) \;\;=\;\;
\overline{\Pi}_{(1)\;\mu\nu}^{\;\;\;\;ab}(x,y) \;\;+\;\;
\overline{\Pi}_{(2)\;\mu\nu}^{\;\;\;\;ab}(x,y)
\label{PiMF}
\end{equation}
\begin{eqnarray}
\overline{\Pi}_{(1)\;\mu\nu}^{\;\;\;\;ab}(x,y)
&=&
\frac{i g}{2}
\;\delta_P^4(x,y)\;
\, \int_P d^4z \,V_{(0)\;\mu\nu\lambda}^{abc}(x,y,z)
\,\overline{A}^{\lambda ,\,c}(z)
\label{PiMF1}
\\
\overline{\Pi}_{(2)\;\mu\nu}^{\;\;\;\;ab}(x,y)
&=&
\frac{g^2}{6}
\;\delta_P^4(x,y)\;
\, \int_P d^4z d^4 w \,W_{(0)\;\mu\nu\lambda\sigma}^{abcd}(x,y,z,w)
\nonumber \\
& & \;\;\;\;\;\;\;\;\;\;\;\;\;\;
\;\;\;\;\;\;\;\;\;\;\;
\times
\,\overline{A}^{\lambda ,\,c}(z) \,\overline{A}^{\sigma ,\,d}(w)
\label{PiMF2}
\;.
\end{eqnarray}
plus terms of order $g^3 \overline{A}^3$ which one may safely ignore
within the present approximation scheme.
The {\it fluctuation polarization tensor} $\widehat{\Pi}$ contains
the quantum self-interaction among the `hard' quanta in the presence of 
$\overline{A}$, and is given by the 
variation of 2-loop part $\Gamma_P^{(2)}$,
of the effective action,
$2i\delta \Gamma_P^{(2)} / \delta\widehat{\Delta}$,
\begin{equation}
\widehat{\Pi}^{ab}_{\mu\nu}(x,y)
\;\;=\;\;
\left(\frac{}{}
\widehat{\Pi}_{(1)}
\;\;+\;\;
\widehat{\Pi}_{(2)}
\;\;+\;\;
\widehat{\Pi}_{(3)}
\;\;+\;\;
\widehat{\Pi}_{(4)}
\right)^{ab}_{\mu\nu}(x,y)
\;,
\label{PiQU}
\end{equation}
\begin{eqnarray}
\widehat{\Pi}^{\;\;\;\;ab}_{(1)\;\mu\nu}(x,y)
&=&
-\, \frac{g^2}{2}
\; \int_P d^4x_1 d^4y_1
\;\;W_{(0)\;\mu\nu\lambda\sigma}^{\;\;\;\;abcd}(x,y,x_1,y_1)\;
\widehat{\Delta}^{\lambda\sigma\,,cd}(y_1,x_1) \;\;\;\;\;\;\;\;
\\
\widehat{\Pi}^{\;\;\;\;ab}_{(2)\;\mu\nu}(x,y)
&=&
-\,\frac{i\,g^2}{2}
\; \int_P\prod_{i=1}^{2} d^4x_id^4y_i 
\;\;V_{(0)\;\mu\lambda\sigma}^{\;\;\;\;acd}(x,x_1,x_2)\;\;
\nonumber \\
& &
\times
\;\widehat{\Delta}^{\lambda\lambda',\,cc'}(x_1,y_1)\;
\widehat{\Delta}^{\sigma\sigma',\,dd'}(x_2,y_2)\;
\widehat{V}_{\sigma'\lambda'\nu}^{d'c'b}(y_2,y_1,y)
\label{Pib}
\\
\widehat{\Pi}^{\;\;\;\;ab}_{(3)\;\mu\nu}(x,y)
&=&
-\;\frac{g^4}{6}
\; \int_P\prod_{i=1}^{3} d^4x_id^4y_i 
\;\;W_{(0)\;\mu\lambda\sigma\tau}^{\;\;\;\;acde}(x,x_1,x_2,x_3)\;\;
\nonumber \\
& &
\times
\;
\widehat{\Delta}^{\lambda\lambda',\,cc'}(x_1,y_1)\;
\widehat{\Delta}^{\sigma\sigma',\,dd'}(x_2,y_2)\;
\nonumber \\
& &
\times
\;
\widehat{\Delta}^{\tau\tau',\,ee'}(x_3,y_3)\;
\widehat{W}_{\tau'\sigma'\lambda'\nu}^{e'd'c'b}(y_3,y_2,y_1,y) 
\label{Pic}
\\
\widehat{\Pi}^{\;\;\;\;ab}_{(4)\;\mu\nu}(x,y)
&=&
-\;\frac{i\,g^4}{24}
\;\int_P\prod_{i=1}^{2} d^4x_id^4y_id^4z_i 
\;\;W_{(0)\;\mu\lambda\sigma\tau}^{\;\;\;\;acde}(x,x_1,x_2,x_3)\;\;
\nonumber \\
& &
\times
\;
\widehat{\Delta}^{\sigma\rho',\,df'}(x_2,z_2)\;
\widehat{\Delta}^{\tau\rho'',\,ef''}(x_3,z_3)\;
\;
\widehat{V}_{\rho''\rho'\rho}^{f''f'f}(z_3,z_2,z_1) \;\;
\nonumber \\
& &
\times
\;
\widehat{\Delta}^{\rho\lambda',\,fc'}(z_1,y_1)\;
\widehat{\Delta}^{\lambda\sigma',\,cd'}(x_1,y_2)\;
\;\;\widehat{V}_{\lambda'\sigma'}^{c'd'}(y_1,y_2,y)
\label{Pid}
\;.
\end{eqnarray}
Note that the usual Dyson-Schwinger equation
in {\it vacuum} is contained in (\ref{DSE2}) -(\ref{Pid}) as the special case 
when the mean field vanishes, $\overline{A}(x)= 0$,
and initial state correlations are absent, ${\cal K}^{(2)}(x,y)=0$. In this case,
the propagator becomes the usual vacuum propagator, 
since the mean-field contribution $\overline{\Pi}$ is identically zero, 
and the quantum part $\widehat{\Pi}$ reduces to the vacuum contribution.

\section{Transition to qantum kinetics}
\label{sec:section4}

The equations of motion (\ref{YME2}) and (\ref{DSE2}) are 
non-linear integro-differential equations and clearly not solvable in
all their generality. To make progress, one must be more specific
and employ now the details of the proclaimed physics scenario, described 
above.

\subsection{Quantum and kinetic space-time regimes}

The key assumption is the separability of hard and soft dynamics in terms 
of the space-time scale $r(\mu)\propto 1/\mu\approx 1$ $fm$.
This implies that one 
may characterize the dynamical evolution of the gluon system
by a short-range {\it quantum scale} $r_{qua}\ll r(\mu)$, and a comparably
long-range {\it kinetic scale} 
$r_{kin}\,\lower3pt\hbox{$\buildrel > \over\sim$}\,r(\mu)$.
Low-momentum collective excitations
that may develop at the particular momentum scale $g\mu$ are
thus well seperated from the typical hard gluon momenta of the order $\mu$,
if $g\ll 1$. Therefore, collectivity can arise, because the wavelength of the
soft oscillations $\sim 1/g\mu$ is much larger than the typical
extention of the hard quantum fluctuations $\sim 1/\mu$.
Notice that this separation of scales is not an academic 
construction, but rather is a general property of quantum field theory.
A simple example is a freely propagating electron:
In this case, the quantum scale is given its the Compton wavelength 
$\sim 1/m_e$ in the restframe of the charge, and measures the size of
the radiative vacuum polarization cloud around the bare charge.
The kinetic scale, on the other hand, is determined by the mean-free-path 
of the charge, which is infinite in vacuum, and in medium is
inversely proportional to the local density times the interaction cross-section,
$\sim 1/(n_g \,\sigma_{int})$.
Adopting this notion
to the present case of gluon  dynamics, let me define
$r_{qua}$ and $r_{kin}$ as follows:

\begin{description}
\item[quantum scale {\boldmath $r_{qua}$}:]
Measures the spatial extension of quantum fluctuations associated with virtual
and real radiative emission and re-absorption
off a given hard gluon, described by the hard propagator $\widehat{\Delta}$. 
It can thus be interpreted as
its Compton wavelength, corresponding
to the typical transverse extension of the fluctuations and
thus inversely proportional to the average transverse momentum,
\begin{equation}
r_{qua} \;\,\equiv\;\,\widehat{\lambda}
\;\;\simeq\;\;\frac{1}{\langle \;k_\perp\;\rangle}
\;,
\;\;\;\;\;\;\;\;\;\;\;\;\;\;\;
\langle \;k_\perp\;\rangle \;\ge \;\mu
\;,
\label{rqua}
\end{equation}
where the second relation is imposed by means of the definition (\ref{Aa})
of hard and soft modes.
Note that $\widehat{\lambda}_C$ is a space-time dependent quantity, because
the magnitude of $\langle k_\perp \rangle$ is determined by both  the
radiative self-interactions of the hard gluons and ther interactions
with the soft field.

\item[kinetic scale {\boldmath $r_{kin}$}:]
Measures the range of the long-wavelength correlations, described by
the soft mean-field $\overline{A}$, and may be parametrized in terms of the
average momentum of soft modes $\langle q \rangle$, such that
\begin{equation}
r_{kin} \;\,\equiv\;\,\overline{\lambda}
\;\;\simeq\;\;\frac{1}{\langle \;q_\perp\;\rangle}
\;,
\;\;\;\;\;\;\;\;\;\;\;\;\;\;\;
\langle \;q_\perp\;\rangle \; \,\lower3pt\hbox{$\buildrel < \over\sim$}\,
\;g\,\mu
\;,
\label{rkin}
\end{equation}
where $\overline{\lambda}$ may vary from one space-time point to
another, because the population of soft modes $\overline{A}(q)$ is determined 
locally by the hard current $\widehat{j}$ with dominant contribution
from gluons with transverse momentum $\simeq \mu$. 
\end{description}

The above classification of quantum- (kinetic-)  scales
specifies in space-time the relevant regime for the hard (soft)
dynamics, and the separability of the two scales 
$r_{qua}$ and  $r_{kin}$
imposes the following condition on the relation between space-time and
momentum: 
\begin{equation}
\widehat{\lambda} \;\;\ll\;\;\overline{\lambda}
\;,\;\;\;\;\;\;\;\;\;\;\;\;\;
\mbox{or}
\;\;\;\;\;\;\;\;\;\;\;\;\;\;
\langle\; k_\perp \;\rangle
\; \;\;\approx
\;\mu\;\;\gg\;\; g\,\mu \;\;
\approx \;\;\;
\langle\; q_\perp \;\rangle
\label{kqcond}
\;.
\end{equation}
The physical interpretation of (\ref{kqcond})
is simple:
At short distances $r_{qua} \ll 1/(g \mu)$ a hard gluon can be considered 
as an {\it incoherent quantum} which emits and partly reabsorbs  other
hard gluons corresponding to the combination of real
bremstrahlung and virtual radiative fluctuatiuons.
Only a hard probe with a short wavelength $\le r_{qua}$ can resolve
this quantum dynamics.
On the other hand,
at larger distances $r_{kin} \approx 1/(g \mu)$, a gluon appears
as a {\it coherent quasi-particle}, that is, as an extended
object with a changing transverse size corresponding
to the extent of its intrinic quantum fluctuations. This dynamical
substructure is however not resolvable by long-wavelength modes
$\ge r_{kin}$ of the soft field $\overline{A}$.
\medskip

Accordingly, one may classify the quantum and kinetic regimes, respectively,
by associating with two distinct space-time points $x^\mu$ and $y^\mu$ the
following characteristic scales:
\begin{eqnarray}
s^\mu 
\;&\equiv&
\;\;
 x^\mu\;-\;y^\mu \;\;
\sim 
\;\widehat{\lambda} \;=\;\frac{1}{g\mu}
\;,\;\;\;\;\;\;\;
\partial_s^\mu 
\;=\;\frac{1}{2}\;\left(\partial_x^\mu\;-\;\partial_y^\mu \right)
\;\;\sim\;\;g\,\mu
\nonumber \\
r^\mu 
\;&\equiv& 
\frac{1}{2}\;\left(x^\mu\;+\;y^\mu\right) 
\;\;\sim 
\;\overline{\lambda} \;=\;\frac{1}{\mu}
\;,\;\;\;\;\;\;\;
\partial_r^\mu 
\;=\; \;\;\;\;\partial_x^\mu\;+\;\partial_y^\mu
\;\;\;\;\sim\;\;\;\mu
\label{sr}
\;.
\end{eqnarray}

The {\it kinetic scale} is therefore $g^2\mu^2$:
The effect of the soft field modes of $\overline{A}$ 
on the hard quanta involves the coupling $g \overline{A}$
to the hard propapgator 
and is of the order of the soft wavelength $\overline{\lambda} = 1/(g\mu)$, 
so that one may characterize the soft field strength by
\begin{equation}
g \overline{A}_\mu(r)  \;\;\sim \;\; g \mu
\;,\;\;\;\;\;\;\;\;\;\;\;\;
g \overline{F}_{\mu\nu}(r)  \;\;\sim \;\;g^2\,\mu^2
\label{AF2}
\;,
\end{equation}
plus corrections of order $g^2\mu^2$ and $g^3\mu^3$, respectively, 
which are assumed to be small.

The {\it quantum scale} on the other hand is $\mu^2$, because 
\begin{equation}
\widehat{\Delta}_{\mu\nu}^{-1}
\;\;\sim k_\perp^2 \;
\,\lower3pt\hbox{$\buildrel > \over\sim$}\,
\;\mu^2
\;\;\gg\;\;g^2\mu^2 \;\; \sim \;\; g\,\overline{F}_\mu\nu
\;,
\end{equation}
and one expects that
that the short-distance fluctuations corresponding to
emission and reabsorption of gluons with momenta $k_\perp \ge \mu$,
are little affected by the long-range, soft mean field, because the
color force $\sim g\overline {F}$ acting on a gluon with momentum
$k_\perp \sim \mu$ produces only a very small change  in its momentum.
\smallskip

Concerning the Yang-Mills equation (\ref{YME2}), 
one finds then immediately from the above scale relations that
both the derivative terms $\partial^2\overline{A}$ and the 
self-coupling terms $\overline{\Xi}$ are of the same order
and need to be included consistently in order to
preserve the gauge symmetry when performing a perturbative
analysis.
Of course,  if the field is weak, $\overline{F}_{\mu\nu}\ll g \mu^2$,
the nonlinear effects contained in the function $\overline{\Xi}$
of (\ref{YME2}) are be subdominant, so that in leading order of $g$,
the color fields would then behave like abelian fields.

\subsection{The kinetic approximation}

The realization of the two space-time scales, short-distance quantum and 
quasi-classical kinetic, allows to reformulate the quantum field theoretical
problem as a relativistic many-body problem within kinetic theory.
The key element is to establish the connection between the preceding
quantum-theoretical description in terms of  Green functions
and a probabilistic kinetic description in terms of
of so-called Wigner functions \cite{wigner}.
Whereas the 2-point functions, such as the propagator or the polarization tensor,
depend on two separate  space-time points $x$ and $y$, 
their Wigner transform 
utilizes a mixed space-time/momentum representation, which is  
particularly convenient
for implementing  the assumption of well separated quantum and
kinetic scales, i.e., that the long-wavelength
field $\overline{A}$ is slowly varying in space-time on the scale
of short-range quantum fluctuations. 
Moreover, the trace of the Wigner transformed propagaor is the quantum analogue 
of  the single particle phase-space distribution of gluons, and
therefore provides the basic quantity to make the connection with 
kinetic theory of  multi-particle dynamics.
\smallskip

In terms of the center-of-mass coordinate,
$r = \frac{1}{2}(x+y)$, and relative coordinate $s=  x-y$,
of two space-time points $x$ and $y$, eq. (\ref{sr}),
one can express any 2-point function 
${\cal G}(x,y)$, such as $\widehat{\Delta},\Pi$, in terms of these coordinates,  
\begin{equation}
{\cal G}_{\mu\nu}^{ab}(x,y)\; =\; {\cal G}_{\mu\nu}^{ab}\left(r+\frac{s}{2}, r-\frac{s}{2}\right)
\;\;\equiv\;\; {\cal G}_{\mu\nu}^{ab}\left( r,s\right)
\;,\;\;\;\;\;\;\;\;\;\;\;\;\;
\;.
\end{equation}
The {\it Wigner transform} ${\cal G}(r,k)$
is then defined as the  Fourier transform with respect to the relative 
coordinate $s$, being the canonical conjugate to the momentum $k$.
In general, the necessary preservation of local gauge symmetry
leads to additional constraint, but for the specific choice
of  gauge (\ref{gauge1}), the Wigner transform is simply 
\begin{equation}
{\cal G}(r,s) \;=\;
\int \frac{d^4k}{(2\pi)^4} \, e^{-i\,k\,\cdot\, s}\;\, 
{\cal G}\left( r,k\right)
\;\;,\;\;\;\;\;\;\;\;\;\;\;\;\;
{\cal G}\left( r,k\right) \;=\;
\int d^4 s \, e^{i\,k\,\cdot\, s}\;\, {\cal G}\left( r,s\right)
\;.
\label{W}
\end{equation}
The Wigner representation (\ref{W}) will facilitate a systematic identification
of the dominant contributions of the soft field $\overline{A}$ to the hard
propagator $\widehat{\Delta}$:
First one expands both $\overline{A}$ and $\widehat{\Delta}_{[\overline{A}]}$
in terms of the long-range variation with the kinetic scale $r$ (gradient expansion),
then one makes an additional expansion in powers of the soft field $\overline{A}$
and of the induced perturbations
$\widehat{\Delta}_{[\overline{A}]} \sim g\widehat{\Delta}_{[\overline{0}]}$.
On this basis, one isolates and
keep consistently terms up to order $g^2\mu^2 \widehat{\Delta}_{[\overline{0}]}$.
\medskip

To proceed, recall that
the coordinate $r^\mu$ describes
the kinetic space-time dependence $O(\Delta r_{kin}$),
whereas $s$ measures the quantum space-time distance $O(\Delta r_{qua}$).
In translational invariant situations, e.g.
in vacuum or thermal equilibrium,  $W(r,s)$ is independent of $r^\mu$ and
sharply peaked about $s^\mu =0$. Here the range of the variation is fixed
by $\widehat{\lambda} = 1/\mu$, eq. (\ref{rqua}), corresponding to
the confinement length $\approx 0.3$ $fm$ in the case of vacuum, 
or to the thermal wavelength $\approx 1/T$ in equilibrium.
On the other hand, in the presence of a slowly varying
soft field $\overline{A}$ with a wavelength  $\overline{\lambda} = 1/(g\mu)$,
eq. (\ref{rkin}),
the $s^\mu$ dependence is little affected, while the acquired $r^\mu$ dependence 
will have a long-wavelength variation.
This suggests therefore to neglect the derivatives of ${\cal G}(r,k)$ with respect to 
$r^\mu$ of order $g\mu$, relative to those with respect to
$s^\mu$ of order $\mu$.

Hence one can perform an expansion of the soft field and the hard propagator and
polarization tensor in terms of gradients,  and keep only terms up to first order,
i.e.,
\begin{eqnarray}
\overline{A}_\mu(x) 
&=&
\overline{A}_\mu\left(r+\frac{s}{2}\right) \;\simeq \;
\overline{A}_\mu(r)+ 
\frac{s}{2}\cdot \partial_r\overline{A}_\mu(r)
\nonumber \\
\overline{A}_\mu(y) 
&=&
\overline{A}_\mu\left(r-\frac{s}{2}\right) \;\simeq \;
\overline{A}_\mu(r)- 
\frac{s}{2}\cdot \partial_r\overline{A}_\mu(r)
\nonumber \\
{\Delta}_{(0)\;\mu\nu}\left(x,y\right)
&=&
{\Delta}_{(0)\;\mu\nu}\left(0,s\right)
\nonumber \\
\widehat{\Delta}_{\mu\nu}\left(x,y\right)
\;\;\;
&=&
\widehat{\Delta}_{\mu\nu}\left(r,s\right) \;\simeq \;
\widehat{\Delta}_{\mu\nu}\left(0,s\right) \;+ \;
s\,\cdot\, \partial_r\, \widehat{\Delta}_{\mu\nu}\left(r,s\right)
\nonumber \\
\overline{\Pi}_{\mu\nu}(x,x)
\;\;\;
&=&
\overline{\Pi}_{\mu\nu}\left(r\right) 
+\frac{s}{2}\cdot \partial_r\overline{\Pi}_{\mu\nu}(r)
\nonumber \\
\widehat{\Pi}_{\mu\nu}\left(x,y\right)
\;\;\;
&=&
\widehat{\Pi}_{\mu\nu}\left(r,s\right) \;\simeq \;
\widehat{\Pi}_{\mu\nu}\left(0,s\right) \;+ \;
s\,\cdot\, \partial_r\, \widehat{\Pi}_{\mu\nu}\left(r,s\right)
\label{gradexp3}
\;,
\end{eqnarray}
and furthermore,
in order to isolate the leading effects of the soft mean field $\overline{A}$
on the hard quantum propagator $\widehat{\Delta}$, one separates
the mean field contribution from the quantum contribution by writing
\begin{equation}
\widehat{\Delta} (r,k)\;\;\equiv\;\;
\widehat{\Delta}_{[\overline{A}]}(r,k) \;\,=\;\, 
\widehat{\Delta}_{[\overline{0}]}(k) \;\,+\;\, 
\delta\widehat{\Delta}_{[\overline{A}]}(r,k)
\label{Dsep}
\end{equation}
with a translation-invariant vacuum quantum contribution and a
$r$-dependent  mean field part, respectively,
\begin{equation}
\widehat{\Delta}_{[\overline{0}]}^{-1} 
\;=\; \left. \widehat{\Delta}^{-1} \right|_{\overline{A}=0}
\;=\;
\Delta_{(0)}^{-1} \;-\; \left.\widehat{\Pi} \right|_{\overline{A}=0}
\;\;\;\;\;\;\;\;\;\;\;\;\;\;\;\;\;\;
\delta\widehat{\Delta}_{[\overline{A}]}^{-1} \;=\;
\overline{\Delta}^{-1} \;-\; \Delta_{(0)}^{-1} 
\;=\; -\; \overline{\Pi}
\label{DA}
\;.
\end{equation}
where  ${\Delta}_{(0)}$ denotes the {\it free}
propagator, and the
$\overline{\Delta}$ the mean-field proagator, that is, the
free propagator in the presence of the mean field, but in the absence
of quantum fluctuations.

Given the ansatz (\ref{Dsep}), with the
feedback of the induced soft field to the hard propagator
being  contained in $\delta\widehat{\Delta}_{[\overline{A}]}$,
one can expand the latter in powers of the soft field coupling $g \overline{A}$, and 
anticipate that it is {\it at most $g$ times} 
the vacuum piece $\widehat{\Delta}_{[\overline{0}]}$,
that is,
\begin{equation}
\delta\widehat{\Delta}_{[\overline{A}]}(r,k) \;=\;
\sum_{n=1,\infty} \frac{1}{n!}\,\left( g\overline{A}(r)\cdot\partial_k\right)^n
\widehat{\Delta}_{[\overline{0}]}(k)
\;\;\simeq\;\;
 g\overline{A}(r)\cdot\partial_k
\widehat{\Delta}_{[\overline{0}]} (k)
\label{DA2}
\;,
\end{equation} 
and, to the same order of approximation, 
$
\partial_r^\mu\delta\widehat{\Delta}_{[\overline{A}]\,\mu\nu}(r,k)
\;\simeq\;
g(\partial_r^\mu\overline{A}^\lambda)\partial_k^\lambda 
\widehat{\Delta}_{[\overline{0}]\;\mu\nu}
$.

Inserting now into eqs. (\ref{YME2}) and (\ref{DSE2}) 
the decomposition (\ref{Dsep}) with the approximation (\ref{DA2}),
and keeping consistently all terms up to order 
$g^2\mu^2 \widehat{\Delta}_{[\overline{0}]}$, one arrives
(after quite some journey \cite{ms42}) at
a set of equations that can be compactly expressed in terms
of the {\it kinetic} momentum $K_\mu$ rather than the {\it canonincal} 
momentum $k_\mu$ (as always in the context of interactions with a
gauge field). For the 
class of gauges gauge (\ref{gauge1a}) amounts to the 
replacements 
\begin{equation}
k_\mu \;\longrightarrow \;K_\mu \;=\; k_\mu \;-\;g \overline{A}_\mu(r)
\;\;,\;\;\;\;\;\;\;\;\;\;
\partial^r_\mu \;\longrightarrow \;
\overline{D}_\mu^r \;=\; \partial^r_\mu \;-\; 
g \partial^r_\mu \overline{A}^\nu(r)\,\partial_\nu^k
\;.
\label{kincan}
\end{equation}
and, within the present approximation scheme,
one has $K^2\widehat{\Delta} \gg  \overline{D}_r^2 \widehat{\Delta}$. 
The result of this procedure is:
\begin{eqnarray}
&&
\;\;
\left[\frac{}{}
\overline{D}_r^{\lambda\;,ab}  , \; \overline{F}_{\lambda \mu}^b
\right](r)
\;\;=\;\;
-\;\widehat{j}_\mu(r) 
\nonumber\\
& & \;\;\;
\;\;=\;\;
- g
\;\int \frac{d^4K}{(2\pi)^2}
\;
\mbox{Tr}\left\{
\;
- K_\mu \, \widehat{\Delta}_{[\overline{A}] \;\nu}^{\;\;\;\;\nu}(r,K)
\;+\;
\widehat{\Delta}_{[\overline{A}] \;\mu}^{\;\;\;\;\nu}(r,K) \,K_\nu
\right\}
\label{YME5}
\;\;\;
\\ & & \nonumber
\\
& &
\;\;
\left\{
\; K^2\, 
,
\; \widehat{\Delta}^{\mu\nu}_{[\overline{0}]}
\right\}
(K)
\;\;=\;\;
\,d^{\mu\nu}(K)\,\,
\;+\;
\frac{1}{2} \;
\left\{\widehat{\Pi}^\mu_{[\overline{0}]\;\sigma}(K)\,,\, 
\widehat{\Delta}^{\sigma\nu}_{[\overline{0}]}(K) \right\}
\label{R2}
\\ & & \nonumber
\\
& &
\;\;
\left[
K\cdot\overline{D}_r , \;  \widehat{\Delta}^{\mu\nu}_{[\overline{A}]}
\;\right]
(r,K)
\;\;=\;\;
-\;\frac{i}{2} \;\left[\overline{\Pi}^\mu_{\sigma}(r,K)\,,\, 
\widehat{\Delta}^{\sigma\nu}_{[\overline{0}]}(K) \right]
\nonumber \\
& & 
\;\;\;\;\;\;\;
\;\;\;\;\;\;\;\;\;\;\;\;
\;\;\;\;\;\;\;\;\;\;\;\;
\;\;\;\;\;\;\;\;\;\;\;\;
\;-\;
\frac{i}{2} \;\left[\widehat{\Pi}^\mu_{[\overline{0}]\;\sigma}(K)\,,\, 
\widehat{\Delta}^{\sigma\nu}_{[\overline{0}]}(K) \right] 
\label{T2}
\;.
\end{eqnarray}
\bigskip

One sees that the original
Dyson-Schwinger equation reduces in the kinetic approximation to
a coupled set of algebraic equations. Recall that
(\ref{R2}) and (\ref{T2})
are still  $2\times 2$ matrix equations mix the four different
components of  $\widehat{\Delta} = (\widehat{\Delta}^F,\widehat{\Delta}^>,
\widehat{\Delta}^<,\widehat{\Delta}^{\overline{F}})$ and of
$\widehat{\Pi} = (\widehat{\Pi}^F,\widehat{\Pi}^>,
\widehat{\Pi}^<,\widehat{\Pi}^{\overline{F}})$.
For the following it is more convenient to
employ instead 
an equivalent  set of independent functions, namely,
the {\it retarded} and {\it advanced functions} $\widehat{\Delta}^{ret}$,
$\widehat{\Delta}^{adv}$, plus 
the {\it correlation function} $\widehat{\Delta}^{cor}$,
and analogously $\widehat{\Pi}$.
This latter set is more directly connected with 
physical, observable quantities, and is commonly referred to as 
{\it physical representation} \cite{chou}:
\begin{equation}
\widehat{\Delta}^{ret} \;=\;   \widehat{\Delta}^F \;-\; \widehat{\Delta}^>
\;\;\;\;\;\;\;\;\;\;\;\;\;\;\;
\widehat{\Delta}^{adv} \;=\; \widehat{\Delta}^F \;-\; \widehat{\Delta}^< 
\;\;\;\;\;\;\;\;\;\;\;\;\;\;\;
\widehat{\Delta}^{cor} \;=\; \widehat{\Delta}^> \;+\; \widehat{\Delta}^< 
\label{retadv1}
\end{equation}
Similarly, for the polarization tensor the retarded, advanced
and correlation functions are defined as 
(note the subtle difference to (\ref{retadv1})):
\begin{equation}
\widehat{\Pi}^{ret} \;=\;   \widehat{\Pi}^F \;+\; \widehat{\Pi}^<
\;\;\;\;\;\;\;\;\;\;\;\;\;\;\;
\widehat{\Pi}^{adv} \;=\; \widehat{\Pi}^F \;+\; \widehat{\Pi}^> 
\;\;\;\;\;\;\;\;\;\;\;\;\;\;\;
\widehat{\Pi}^{cor} \;=\; -\widehat{\Pi}^> \;-\; \widehat{\Pi}^< 
\label{retadv2}
\end{equation}
Loosely speaking, the retarded and advanced functions characterize
the intrinsic quantum nature of 
a `dressed' gluon, describing its substructural state of 
emitted and reabsorbed gluons, whereas the correlation function describes
the kinetic correlations among different such `dressed' gluons.
The great advantage \cite {chou} of this  physical representation is that 
in general the dependence 
on the phase-space occupation of gluon states (the local density) 
is essentially carried by the correlation functions $\widehat{\Delta}^{>}$,
$\widehat{\Delta}^<$, 
whereas the dependence of the retarded and advanced functions, 
$\widehat{\Delta}^{ret}$, $\widehat{\Delta}^{adv}$,
on the local density is weak.
More precisely,
the retarded and advanced propagators
and the imaginary parts of the self-energies embody the
renormalization effects and dissipative quantum dynamics that is associated
with short-distance emission and absorption of quantum fluctuations,
whereas the correlation function contains both the effect of interactions
with the soft mean  field and of statistical binary scatterings among the hard
gluons.
In going over to the physical representation,  one arrives at the
set of master equations:
The Yang-Mills equation (\ref{YME5}) reads
\begin{eqnarray}
&&
\;\;
\left[\frac{}{}
\overline{D}_r^{\lambda}  , \; \overline{F}_{\lambda \mu}
\right]
\;\;=\;\;
- g
\;\int \frac{d^4k}{(2\pi)^2}
\;
\mbox{Tr}\left\{
\;
\left(
- K_\mu \,\widehat{\Delta}_{[\overline{A}]\;\nu}^{cor\;\nu}
\;+\;
\widehat{\Delta}_{[\overline{A}]\mu}^{cor\;\nu} \,K_\nu
\right) 
\right\}
\label{YME6}
\;\;\;\;
\end{eqnarray}
and
the renormalization (\ref{R2})  and transport equations (\ref{T2}),
become \cite{ms39}
\begin{eqnarray}
& &
\left\{
\;K^2\,, 
\; \widehat{\Delta}^{ret}_{[\overline{0}]}-\widehat{\Delta}^{adv}_{[\overline{0}]}
\right\}_{\mu\nu}
\;=\;
-\frac{1}{2}
\left\{{\cal M}^2\, , \,  \mbox{Im}  \widehat{\Delta}_{[\overline{0}]}
\right\}_{\mu\nu}
\;-\;
\frac{1}{2}
\left\{ \Gamma\, , \,  \mbox{Re}\widehat{\Delta}_{[\overline{0}]}
\right\}_{\mu\nu}
\label{X1}
\;\;\;\;\;\;
\\
& \nonumber \\
& &
\left[
K\cdot \overline{D}_r  \, ,\,  \widehat{\Delta}^{cor}_{[\overline{A}]}
\right]_{\mu\nu}
\;=\;
+\frac{i}{2}
\left[ {\Pi}^{cor}\, , \,  \mbox{Re} \widehat{\Delta}_{[\overline{0}]}
\right]_{\mu\nu}
\;-\;
\frac{1}{4} 
\left\{ {\Pi}^{cor}\, , \,  \mbox{Im}\widehat{\Delta}_{[\overline{0}]}
\right\}_{\mu\nu}
\nonumber \\
& &
\;\;\;\;\;\;\;\;\;\;\;\;\;\;
\;\;\;\;\;\;\;\;\;\;\;\;\;\;\;\;\;\;\;\;\;
\;+\;
\frac{i}{2}
\left[ {\cal M}^2\, , \,  \widehat{\Delta}^{cor}_{[\overline{0}]}
\right]_{\mu\nu}
\;-\;
\frac{1}{4} 
\left\{ \Gamma\, , \,  \widehat{\Delta}^{cor}_{[\overline{0}]}
\right\}_{\mu\nu}
\;,
\label{X2}
\end{eqnarray}
where $\,\Pi \,=\,\overline{\Pi}\,+\,\widehat{\Pi}$,
and  the real and imaginary components of the 
polarization tensor are denoted by
\begin{equation}
{\cal M}^2_{\mu\nu} \;\equiv\;
\mbox{Re} {\Pi}_{\mu\nu}\;=\; \frac{1}{2} \,\left( {\Pi}^{ret}+{\Pi}^{adv}\right)_{\mu\nu}
\;\;\;\;\;\;
\Gamma_{\mu\nu} \;\equiv\;
\mbox{Im} {\Pi}_{\mu\nu}\;=\; i \, \left( {\Pi}^{ret}-{\Pi}^{adv}\right)_{\mu\nu}
\end{equation}
Note also, that 
are the real and imaginary components of the 
Hard propagator are given by the sum and difference of the retarded
and advanced contributions, respectively,
\begin{equation}
\mbox{Re} \widehat{\Delta}_{\mu\nu}\;=\; 
\frac{1}{2}\, \left( \widehat{\Delta}^{ret}+\widehat{\Delta}^{adv}\right)_{\mu\nu}
\;\;\;\;\;\;\;\;\;\;\;\;
\mbox{Im} \widehat{\Delta}\;=\; i \,
\left( \widehat{\Delta}^{ret}-\widehat{\Delta}^{adv}\right)_{\mu\nu}
\;.
\end{equation}
The physical significance of the (\ref{X1}) and (\ref{X2}) is
the following:
Eq. (\ref{X1}) determines the state of a dressed parton with
respect to their virtual fluctuations and real emission (absorption) processes,
corresponding to the real and imaginary parts of the retarded and advanced
self-energies.
Eq. (\ref{X2}), on the other hand characterizes the correlations
mong different dressed parton states, 
and the self-energies appear here in two distinct ways.
The first two terms on the right hand side account for scatterings between quasi-particle
states, i.e. dressed partons, whereas the last two terms incorporate the renormalization effects 
which result from the fact that the dressed partons between collisions do not behave as
free particles, but change their dynamical structure due to virtual
fluctuations, as well as real emission and absorption of quanta.
For this reason ${\Pi}^\ra$ is called {\it radiative} self-energy, and ${\Pi}^{cor}$
is termed {\it collisional} self-energy.
It is well known \cite{chou}, that the imaginary parts of
the retarded and advanced Green functions  and self-energies are
just the spectral density $\rho = \mbox{Im}\widehat{\Delta}$, 
giving the probability for finding an intermediate
multi-particle state in the dressed parton, respectively the decay width $\Gamma$,
describing the dissipation of the dressed parton. 
The formal solution of  (\ref{X1}) and (\ref{X2}) for
the spectral density ${\rho}$ is 
\begin{equation}
{\rho}(r,k)\;=\;
\frac{\Gamma}{k^2 \,-\,{\cal M}^2\,+\,(\Gamma/2)^2}
\;\,\equiv\;\,
\,{\rho}_{{\cal M}^2}
\;+\; {\rho}_{\Gamma}
\;,
\label{X4}
\end{equation}
describing the particle density in terms of the finite width $\Gamma$
and the dynamical `mass term' ${\cal M}^2$ (which in the `free-field'
case are $\Gamma = {\cal M}^2= 0$, corresponding to
an on-shell, classically stable particle).
On the right hand side of (\ref{X4}), the second form 
exhibits the physical meaning more suggestively in
terms of the `wavefunction'-renormalization
(${\rho}_{{\cal M}^2}={\rho}_{\Gamma =0}$) due to
virtual fluctuations, and 
the dissipative parts (${\rho}_{\Gamma}=\rho_{{\cal M}^2=0}$)
due to real emission (absorption) processes.

\section{Outlook}

What remains to be done is to solve the set of equations (\ref{YME6})-(\ref{X2})
which is the hardest part. For the case of 
$\overline{A}_\mu = 0  = \overline{F}_{\mu\nu}$, this was discussed in
Ref. \cite{ms39}. For the present general case, the coupling between
hard gluons ($\widehat{\Delta}$) and the soft field ($\overline{A}$), complicates
things considerably.
A possible iterative scheme of solution, which is currently under
study \cite{ms42}, may be as follows:
\begin{description}
\item[a)]
Specify initial condition in terms of a phase-space density
of hard gluons at time $t=t_0$. 
This initial gluon distribution determines
$\widehat{\Delta}_{[\overline{0}]}(t=t_0,\vec{r}, k)$.
\item[b)]
Solve the renomalization equation (\ref{R2}) with $\overline{A}(t_0,\vec{r}) = 0$,
i.e. just as in the case of vacuum \cite{ms39}, except that now
$K = k-g\overline{A}$ contains the soft field.
Substitute the resulting form of
$\widehat{\Delta}^{ret}_{[\overline{0}]}$ and $\widehat{\Delta}^{adv}_{[\overline{0}]}$ into the transport equation (\ref{T2}) to get the solution
for $\widehat{\Delta}^{cor}_{[\overline{0}]}$.
\item[c)] 
Insert 
$\widehat{\Delta}^{cor}_{[\overline{0}]}$ into the right hand side of the Yang-Mills
equation (\ref{YME6}), and solve for $\overline{A}$.
\item[d)]
Repeat from a) but now include the finite contribution from the
coupling between
$\widehat{\Delta}_{[\overline{0}]}$ and $\overline{A}$.
\end{description}


\newpage

\end{document}